\documentclass[aps, showpacs, showkeys,nofootinbib,floatfix]{revtex4}

\usepackage{amssymb}
\usepackage{amsmath}
\usepackage{graphicx}
\usepackage{hyperref}

\begin{document}

\title{The Dynamic Behavior of Quantum Statistical Entropy in 5D Ricci-flat Black String with Thin-layer Approach}

\author{Molin Liu}
\email{mlliudl@student.dlut.edu.cn}
\author{Hongya Liu}
\email{hyliu@dlut.edu.cn}

\affiliation{School of Physics and Optoelectronic Technology,
Dalian University of Technology, Dalian, 116024, P. R. China}

\begin{abstract}
In this paper, the statistical-mechanical entropies of 5D
Ricci-flat black string is calculated through the wave modes of
the quantum field with improved thin-layer brick-wall method. The
modes along the fifth dimension are semi-classically quantized by
Randall-Sundrum mass relationship. We use the two-dimensional area
to describe this black string's entropy which, in the small-mass
approximation, is a linear sum of the area of the black hole
horizon and the cosmological horizon. The proportionality
coefficients of entropy are discretized with quantized extra
dimensional modes. It should be noted that the small-mass
approximation used in our calculation is naturally justified by
the assumption that the two branes are located far apart.
\end{abstract}



\maketitle

\section{Introduction}
In 1970s, Hawking \cite{Hawking1} justified that the black holes
should be treated as a thermodynamic system which contributes the
temperature associated with its surface gravity, viz
\begin{equation}\label{temperature}
    T = \frac{\kappa}{2 \pi},
\end{equation}
where $\kappa$ is the surface gravity at the horizon. Shortly
before this discovery, the concept of black hole entropy is
originally developed by Bekenstein \cite{Bekenstain}. The famous
result is that the entropy of black hole is proportional to the
area of its horizon. The quantitative entropy is
\begin{equation}\label{entropy0}
    S = \frac{1}{4} A,
\end{equation}
where $A$ is the area of horizon. In this way, there are many
literatures \cite{Howtounderstand} studying the origin of black
hole entropy by various approaches. One of them is the famous
brick wall method (BWM) shown by 't Hooft. In this model, there is
a brick wall around event horizon. In order to avoid the
occurrence of divergent entropy (or free energy), the matter field
is assumed to be vanished beyond a mini-distance outside black
hole. Then brick wall and black hole construct a thermodynamic
system. The entropy of black hole is identified with the
statistical mechanical entropy created by the excitation of
external quantum field. However, the application of BWM must
satisfy the condition that the external field should be in thermal
equilibrium with black hole in a large spatial region, i.e., BWM
should not be used directly to the nonequilibrium system such as a
black hole with multi-horizon. Recently, this confinement was
solved by an improved BWM --- the thin-layer approach \cite{Li}.
In the thin-layer model, the large thermodynamics equilibrium is
replaced by the local equilibrium on a microscopic scale. The
mathematical difference of integration range implies a distinct
physical significance. As an effective approach to nonequilibrium
systems, the thin-layer model is usually employed to study many
types of multi-horizon spaces, for instance, Schwarzschild-de
Sitter black hole\cite{SdS}, Kerr-de Sitter black hole \cite{KdS},
Vaidya black hole \cite{Li} and so on \cite{others}.  In this
paper, we try to use the thin-layer method to study the entropy of
a 5D nonequilibrium black string, which there are two horizons
that one is event horizon and the other is cosmological horizon.

Here the Space-Time-Matter (STM) \cite{ref:Wesson}
\cite{ref:Overduin} theory is mentioned as the foundation of this
work. In STM, the extra dimension is non-compacted. The 5D
manifold is Ricci-flat while the 4D hypersurface is curved by the
4D induced matter. Mathematically, this approach is supported by
Campbell's theorem \cite{ref:Campbell}. Recently an accelerating
universe has proposed to be a way to interpret the astronomical
data of type Ia supernovae \cite{Perlmutter}. Combining this
astonishing observation in the standard cosmology leads to that
our universe approaches de Sitter geometries in both past and
future \cite{Witten}. Hence we consider a Schwarzschild-de Sitter
black hole embedded into a 5D Ricci-flat space. A class of 5D
Ricci-flat black hole solutions containing a 4D de Sitter space
are shown by Mashhoon et al \cite{ref:Wesson} \cite{ref:Mashhoon}
\cite{ref:Liu}. In our previous work \cite{ref:Liu00}, a system
with double Randall-Sundrum (RS) branes
 is constructed using these solutions. If matter trapped on the brane undergoes gravitational
collapse, a black hole will form naturally and its horizon extends
into the extra dimension which is transverse to brane. Such higher
dimensional object looked like a black hole on the brane is
actually a black string in the higher dimensional brane world.
Hence a 5D Ricci-flat black string is obtained naturally. As one
candidate of higher dimensional black holes, its thermodynamical
aspect is needed to be studied.

On the other hand, in the last decade of 20th century the
additional spacelike dimensional ADD model \cite{ADD} and RS model
\cite{RS1} \cite{RS2} raise the upsurge of study higher
dimensional brane world. Based on these brane world model, various
black holes are studied \cite{Dblackhole} and their entropy also
becomes an interesting topic \cite{Dentropy}, and the STM theory
is equivalent to the brane world model \cite{Ponce} \cite{Seahra}
\cite{Liu_plb} \cite{Ping}. Therefore it is worth to study the
entropy of 5D Ricci-flat black string with a 4D effective
cosmological constant.

This paper is organized as follows: In Section II, the 5D
Ricci-flat black string metric and the surface gravity near
horizons are presented. In section III, by semi-classical method
the quantized modes are obtained with WKB approximation. In
section VI, the entropy of 5D Ricci-flat black string is
calculated with the assumption of far apart two branes. We adopt
the signature $(+, -, -, -, -)$ and put $\hbar$, $C$ ,and $G$
equal to unity. Greek indices $\mu, \nu, \ldots$ will be taken to
run over $0, 1, 2, 3$ as usual, while capital indices A, B, C,
$\ldots$ run over all five coordinates $(0, 1, 2, 3, 4)$.
\section{5D Ricci-flat black string with an effective cosmological constant}
A static, three-dimensional spherically symmetric 5D line element
having a 4D effective cosmological constant takes the form
\cite{ref:Wesson} \cite{ref:Mashhoon} \cite{ref:Liu}
\begin{equation}
d s^{2}=\frac{\Lambda
\xi^2}{3}\left[f(r)dt^{2}-\frac{1}{f(r)}dr^{2}-r^{2}\left(d\theta^2+\sin^2\theta d\phi^2\right)\right]-d\xi^{2}, \label{eq:5dmetric}%
\end{equation}
which actually is a black brane solution in brane world. The
metric function is
\begin{equation}
f(r)=1-\frac{2M}{r}-\frac{\Lambda}{3}r^2,\label{f-function}
\end{equation}
where $\xi$ is an open non-compact extra dimension coordinate and
$M$ is the central mass. One should note that the above $\Lambda$
is an induced cosmological constant which is obtained by the
reduction from 5D to 4D. In another words, since this metric
(\ref{eq:5dmetric}) is Ricci-flat $R_{AB}=0$, there is no
cosmological constant in 5D space. So one can actually deal with
this $\Lambda$ as a parameter which comes from the fifth
dimension. The part of this metric inside the square bracket is
exactly the same line-element as the 4D Schwarzschild-de Sitter
solution, which is bounded by two horizons
--- an inner horizon (event horizon) and an outer horizon
(cosmological horizon). This solution has been studied in many
works \cite{ref:Mashhoon11} focusing mainly on the induced
constant $\Lambda$, the extra force and so on.

In the work of \cite{ref:Liu00}, the binary Randall-Sundrum branes
system is constructed. Now we briefly list the nontrivial results
below. By a coordinate transformation $\xi=\sqrt{3 / \Lambda}\
\text{exp}(\sqrt{\Lambda / 3}\ y)$ \cite{ref:Liu00}, the metric
(\ref{eq:5dmetric}) then takes a manifestly conformally form
\begin{equation}
d s^{2}=e^{2\sqrt{\frac{\Lambda}{3}}y}\left[f(r)dt^{2}-\frac{1}{f(r)}dr^{2}-r^{2}\left(d\theta^2+\sin^2\theta d\phi^2\right)-dy^{2}\right].\label{eq:5dmetric-y}%
\end{equation}
Then we use this metric to construct a RS type brane model in
which the first brane is at $y=0$, and the second brane is at
$y=y_{1}$. In this way the double brane model is obtained, and the
fifth dimension becomes finite. It could be very small as in RS
2-brane model \cite{RS1} or very large as in RS 1-brane model
\cite{RS2}, and it is a black string intersecting the brane world,
that is to say, on the hypersurface of fixed extra dimension this
metric describes a SdS black hole. However, when viewed from 5D,
the horizon does not form a 4D sphere --- it looks like a black
string lying along the fifth dimension. Hence we call the solution
(\ref{eq:5dmetric}) black string. Arnowitt-Deser-Misner (ADM) mass
$\widetilde{M}$ of the black string measured on the second brane
is
\begin{equation}
\widetilde{M} = M e^{\sqrt{\Lambda/3}\, y_1},
\label{RSrelationofmass}
\end{equation}
while its ADM mass $\widetilde{M} = M$ locating at the first
brane.

The metric function (\ref{f-function}) can be recomposed as
follows
\begin{equation}
f(r)=\frac{\Lambda}{3r}(r-r_{e})(r_{c}-r)(r-r_{o}). \label{re-f function}%
\end{equation}

The singularity of the metric (\ref{eq:5dmetric-y}) is determined
by $f(r)=0$. Here we only consider the real solutions. The
solutions to this equation are black hole event horizon $r_{e}$,
cosmological horizon $r_{c}$ and a negative solution
$r_{o}=-(r_{e}+r_{c})$. The last one has no physical significance,
and $r_{c}$ and $r_{e}$ are given as

\begin{equation}
\left\{
\begin{array}{c}
r_{c} = \frac{2}{\sqrt{\Lambda}}\cos\chi ,\\
r_{e} = \frac{2}{\sqrt{\Lambda}}\cos(\frac{2\pi}{3}-\chi),\\
\end{array}
\right.\label{re-rc}
\end{equation}
where $\chi=\frac{1}{3}\arccos(-3M\sqrt{\Lambda})$ with $\pi/6
\leq\chi\leq \pi/3$. The real physical solutions are accepted only
if
 $\Lambda$ satisfy $\Lambda M^2\leq\frac{1}{9}$ \cite{ref:Liu}.

So according to the metric function, the tortoise coordinate
is
\begin{equation}
x=\frac{1}{2M}\int\frac{dr}{f(r)}.\label{tortoise }
\end{equation}
Integration of this equation shows that $x$ can be expressed explicitly in the following form:%
\begin{equation}
x=\frac{1}{2M}\left[\frac{1}{2K_{e}}\ln\left(\frac{r}{r_{e}}-1\right)-\frac{1}{2K_{c}}\ln\left(1-\frac{r}{r_{c}}\right)+\frac{1}{2
K_{o}}\ln\left(1-\frac{r}{r_{o}}\right)\right],\label{tortoise of
gravitaion surface}
\end{equation}
where%
\begin{equation}
K_{i}=\frac{1}{2}\left|\frac{df}{dr}\right|_{r=r_i}.
\end{equation}
That is
\begin{equation}
    K_{e}=\frac{(r_{c}-r_{e})(r_{e}-r_{o})}{6r_{e}}\Lambda,\label{Ge}
\end{equation}
\begin{equation}
    K_{c}=\frac{(r_{c}-r_{e})(r_{c}-r_{o})}{6r_{c}}\Lambda,\label{Gc}
\end{equation}
\begin{equation}
   K_{o}=\frac{(r_{o}-r_{e})(r_{c}-r_{o})}{6r_{o}}\Lambda.\label{Go}
\end{equation}
According to Hawking temperature formula (\ref{temperature}),
there are two distinct temperatures near event horizon $r_{e}$ and
cosmological horizon $r_{c}$. So the 5D black string space is a
nonequilibrium system apparently. Therefore, the BWM can not be
applied directly to this multi-horizon space. So the proper
selection is thin-layer approach --- improved BWM method.

Being a minimally coupled quantum scalar field $\Phi$ with mass
$m_{0}$, the field equation on the background
(\ref{eq:5dmetric-y}) is
\begin{equation}
\frac{1}{\sqrt{g}}\frac{\partial}{\partial
x^{A}}\left(\sqrt{g}g^{AB}\frac{\partial}{\partial{x^{B}}}\right)\Phi - m_{0}^2\Phi = 0.\label{Klein-Gorden equation}%
\end{equation}
The modes of the scalar field can be decomposed as the separable
form,
\begin{equation}
\Phi = e^{-iEt} R_{\omega}(r)L(y)Y_{lm}(\theta,\phi),\label{wave
function}
\end{equation}
where $E$ is the particle energy and  $Y_{lm}(\theta,\phi)$ is the
usual spherical harmonic function. Then the equations for $L(y)$
and $R_{\omega}(r)$ read as follows,

\begin{eqnarray}
      && e^{-3\sqrt{\frac{\Lambda}{3}}y}\frac{d}{d y}\left (e^{3\sqrt{\frac{\Lambda}{3}}y}\frac{d}{d
    y}\right )L(y) + \left (e^{2\sqrt{\frac{\Lambda}{3}}y}m_{0}^2 + \mu^2\right
    )L(y) = 0;\label{L} \\
   && E^2 \frac{1}{f(r)} R_{\omega}(r) + \frac{1}{r^2}\frac{d}{dr}\left(r^2 f(r) \frac{d}{d
   r}\right)R_{\omega}(r) - \left (\mu^2 + \frac{l(l+1)}{r^2}\right
   ) R_{\omega} (r) = 0. \label{r}
\end{eqnarray}
The eigenvalue $\mu^2$ is the effective mass on the brane and
Eq.(\ref{r}) is exactly the same as the usual radial equation of
massive scalar particle around 4D SdS black hole. The similar
effective mass can be found in the entropy of RS black string
\cite{Dentropy}. According to the above ADM mass relationship
 (\ref{RSrelationofmass}), one can get the effective mass $\mu$ located on the two
 branes,
\begin{equation}
\mu =\left\{
\begin{array}{c}
 m_0,{\ \ \ \ \ \ \ \ \ \ \ \ \ \ \ \ \ }\text{\ the\ first\ brane};\\
m_0 e^{\sqrt{\Lambda/3}\,y_1},{\ \ \ \ \ \ \ }\text{\ the\ second\ brane}.\\
\end{array}
\right.\label{mumass}
\end{equation}

\section{Semi-classical Quantized Modes Along Extra Dimension}
In the interest of the entropy of this black string, we
investigate the modes along the fifth dimension $y$ before the
other four dimensions ($t,\ r,\ \theta,\ \phi$). It is expected to
find the definite expression of $\mu$ which plays a role of the
effective mass on the brane. So Eq. (\ref{L}) is simplified as
follows,
\begin{equation}\label{L2}
    \frac{d^2 L(y)}{d y^2} + \sqrt{3\Lambda} \frac{d L(y)}{d y} +
    \left(e^{2\sqrt{\frac{\Lambda}{3}}\, y}m_{0}^{2} +
    \mu^2\right)L(y)=0.
\end{equation}
We assume the modes of the fifth dimension is $L(y) = e^{iY(y)}$
and the wave number $k_y$ of wave function $L(y)$ is $k^2_y =
\left(\frac{\partial Y}{\partial y}\right)^2$. Here we employ
Wentzel-Kramers-Brillouin (WKB) approximation \cite{Hooft} in
which the wave function is expanded in a series of Planck constant
$\hbar$ (small quantity). So the result can be obtained by
stagewise approximation. By WKB approximation \cite{Hooft}, the
wave number $k_y$ can be written as
\begin{equation}\label{wavenumbery}
    k_y^2 = e^{2\sqrt{\frac{\Lambda}{3}}\ y} m_0^2 + \mu^2.
\end{equation}

\begin{figure}
  \includegraphics[width=3.5 in]{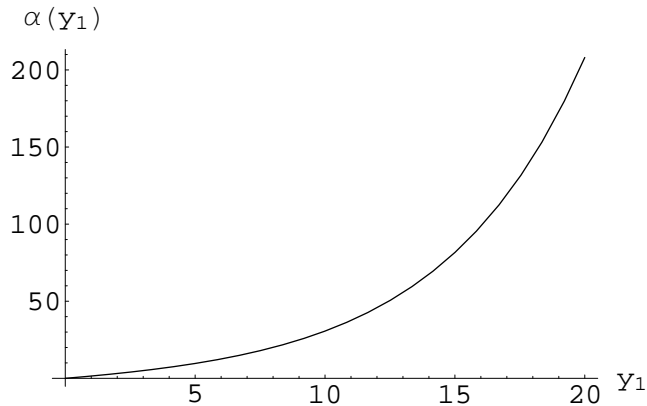}\\
  \caption{The ADM mass factor versus the brane position $y_1$ ($\Lambda = 0.1$ is assumed).}\label{fig1}
\end{figure}

The extra dimension wave number $n_y$ satisfies the semi-classical
quantization condition
\begin{eqnarray}\label{ny}
   \nonumber \pi n_y (\mu) &=& \int_0^{y_1} dy\ k_y (y, \mu)\\
   \nonumber &=&\sqrt{\frac{3}{\Lambda}} \Bigg{\{}-\sqrt{(m_0 -\mu)(m_0 +\mu)} + \sqrt{e^{2 \sqrt{\frac{\Lambda}{3}}
   y_1}m_0^2-\mu^2}\\
   &+& \mu \cot^{-1}\bigg{(}\frac{\mu}{\sqrt{m_0^2-\mu^2}}\bigg{)}-\mu
   \cot^{-1}\left(\frac{\mu}{\sqrt{e^{2\sqrt{\frac{\Lambda}{3}\,y_1}-1}}}\right)\Bigg{\}},
\end{eqnarray}
where the tensions of branes are not restricted. It is very hard
to directly solve $\mu$ from the above function. In order to
simplify calculation, the effective ADM mass relationship
(\ref{mumass}) is adopted here. If the first brane $y = 0$ is our
real world, the RS relation is reduced to $\mu = m_0$.
Substituting this ADM mass relationship into the quantization
condition (\ref{ny}), the expected expression of effective ADM
mass can be obtained as
\begin{equation}\label{muexpress}
    \mu = \frac{\pi n_y}{\alpha}\ \ \ (n_y = 1, 2, 3, \ldots),
\end{equation}
where
\begin{equation}\label{alpha}
    \alpha = \sqrt{e^{2\sqrt{\frac{\Lambda}{3}}y_1}-1} -
    \cot^{-1}\left(\frac{1}{\sqrt{e^{2
    \sqrt{\frac{\Lambda}{3}}y_1}-1}}\right).
\end{equation}
So the modes of extra dimension is quantized and the mass is
discretized naturally. Apparently, $\alpha$ can be used to
determine the relationship between ADM mass and the position of
brane. We call parameter $\alpha$ the ADM mass factor whose
function is drawn in Fig. \ref{fig1}.
It is clear that if $y_1 \longrightarrow \infty$, the limit of
$\alpha \longrightarrow \infty$ exists and the modes spectrum is
continuous. Similar behavior can be found in the work of the RS
black string \cite{Dentropy}.

\section{Entropy of the Ricci-flat black string}
We assume the radial wave function satisfies $R_{\omega}(r) \sim
\exp(i S(r))$. Making use of the WKB approximation, we can obtain
a r-dependent radial wave number $k_{r}(r)$ by
\begin{equation}\label{rwavenumber}
   k_r^2 = \left(\frac{\partial S(r)}{\partial r}\right)^2 =
   \left[E^2 f^{-1}(r) - \left(\mu^2 +
   \frac{l(l+1)}{r^2}\right)\right]f^{-1}(r).
\end{equation}
In usual 4D space, the mass $\mu$ is always treated as a small
mass approximation and sometimes is ignored during calculation.
However, it is not necessary to do this and the $\mu$ is kept as
the effective ADM mass in higher dimensional gravity such as in
Randall-Sundrum black string \cite{Dentropy}. The number of radial
wave $n_r$ is also obtained via the semi-classical quantized
condition,
\begin{equation}\label{wavenumer}
    \pi n_r = \int _r k_r (r) dr.
\end{equation}
So the total number of modes $N_r$ with energy less than or equal
to $E$ is given by
\begin{equation}\label{N}
    \pi N_r = \int (2l+1)\pi n_r dl.
\end{equation}
According to the canonical assembly theory with definite particles
number $N$, volume $V$ and temperature $T$, the free energy of the
quantum scalar field at inverse temperature $\beta$ is written as
\begin{eqnarray}\label{freeenergy}
    \nonumber\beta F &=& \sum_{E} \left(1 - e^{-\beta E}\right)\\
&=& \int _E g(E) \ln\left(1 - e^{-\beta E}\right) d E = \int
_0^{+\infty} \ln\left(1 - e^{-\beta E}\right) d N_r,
\end{eqnarray}
where $g(E) = d N_r (E)/ d E$ is the density of states and the
summation is substituted by integration with semiclassical view.
Substituting state function $N_r(E)$ into Eq. (\ref{freeenergy}),
we get
\begin{equation}\label{freeenergy1}
    F = -\frac{1}{\pi}\int_{0}^{+\infty}d E \int_r
d r \int _l (2l+1)\ d l\ \frac{k_r(r,\ E,\ l)}{e^{\beta E}-1}.
\end{equation}
The $l$ integral can be simplified by restricting integral range,
and the F can be explicitly written as
\begin{equation}\label{freeenergy2}
    F = -\frac{2}{3\pi}\int_0^{+\infty}\frac{d E}{e^{\beta
    E}-1}\int_r d r\ \frac{r^2}{f^2 (r)}\ \left[E^2 - f(r)
    \mu^2\right]^{3/2}.
\end{equation}
In order to separate the variable $\mu$ and metric function
$f(r)$, the integrand can be expanded in a series. Here we assume
the two branes are very far apart each other. A new parameter
$\epsilon = \mu^2 = \pi^2 n_y^2/\alpha^2$ is introduced. When the
second brane is sent to infinite, i.e. $\alpha \longrightarrow
\infty$, the parameter $\epsilon$ can be treated naturally as
small quantity according to the effective ADM mass
(\ref{muexpress}). So using the relationship $(E^2-f(r)
\epsilon)^{3/2} \approx E^3 -3/2 f(r) E \epsilon$, the free energy
(\ref{freeenergy2}) is rewritten as
\begin{eqnarray}\label{freeenergy3}
   \nonumber F &=& -\frac{2}{3\pi}\int_r d r \int_0^{+\infty}
    \frac{r^2}{f^2(r)} \frac{E^3}{e^{\beta E}-1} d E +
    \frac{1}{\pi} \int_r d r \int_0^{+\infty} \frac{r^2
    \epsilon}{f(r)}\frac{E}{e^{\beta E}-1}d E\\
&=& -\frac{2\pi^3}{45 \beta^4}\int_r\frac{r^2}{f^2(r)}d r +
\frac{\pi}{6 \beta^2} \int_r \frac{r^2\epsilon}{f(r)} dr.
\end{eqnarray}
The first term is exactly the same as the usual 4D SdS case
\cite{wenbiao} and the second shows the effect of extra dimension
in brane world. Then the integral range of radial direction is
determined by the improved thin-layer BWM boundary conditions
\begin{eqnarray}
  \Phi (t,\ r,\ \theta,\ \phi,\ y) &=& 0\ \ \ \ \text{for}\ \  r_e + \varepsilon_e \leqslant r \leqslant r_e + \varepsilon_e + \delta_e;\\
  \Phi (t,\ r,\ \theta,\ \phi,\ y) &=& 0\ \ \ \ \text{for}\ \  r_c -
  \varepsilon_c - \delta_c \leqslant r \leqslant r_c -
  \varepsilon_c,
\end{eqnarray}
where $\varepsilon_e$ and $\varepsilon_c$ are the infinitesimal
cutoff factors ($\varepsilon_e,\ \varepsilon_c \ll \ r_e\
\text{or}\ r_c$); $\delta_e$ and $\delta_c$ are the thickness of
thin-layer. Then the main contributions of the integration near
horizons $r_e$ and $r_c$ are given respectively by
\begin{eqnarray}
  \nonumber F_e &=& -\frac{2\pi^3}{45 \beta^4}\int_{r_e + \varepsilon_e}^{r_e + \varepsilon_e + \delta_e}\frac{r^2}{f^2(r)}d r +
\frac{\pi}{6 \beta^2} \int_{r_e + \varepsilon_e}^{r_e + \varepsilon_e + \delta_e} \frac{r^2\epsilon}{f(r)} dr \\
&=&-\frac{2\pi^3}{5\beta^4\Lambda^2}\cdot\frac{r_e^4}{(r_e-r_o)^2(r_e-r_c)^2}\frac{\delta_e}{\varepsilon_e(\varepsilon_e
+
\delta_e)}+\frac{\pi\epsilon}{6\beta^2}\cdot\frac{r_e^2}{(r_e-r_o)(r_c-r_e)}\ln\frac{\varepsilon_e
+ \delta_e}{\varepsilon_e};\\
\nonumber  F_c &=& -\frac{2\pi^3}{45 \beta^4}\int_{r_c -
\varepsilon_c - \delta_c}^{r_c - \varepsilon_c}\frac{r^2}{f^2(r)}d
r + \frac{\pi}{6 \beta^2} \int_{r_c - \varepsilon_c -
\delta_c}^{r_c - \varepsilon_c} \frac{r^2\epsilon}{f(r)} dr\\
&=&-\frac{2\pi^3}{5\beta^4\Lambda^2}\cdot\frac{r_c^4}{(r_c-r_o)^2(r_c-r_e)^2}\frac{\delta_c}{\varepsilon_c(\varepsilon_c
+
\delta_c)}+\frac{\pi\epsilon}{6\beta^2}\cdot\frac{r_c^2}{(r_c-r_o)(r_c-r_e)}\ln\frac{\varepsilon_c}{\varepsilon_c
+ \delta_c}.
\end{eqnarray}
Here the total free energy is $F = F_e + F_c$. Hence, the entropy
of 5D Ricci-flat black string is given by
\begin{eqnarray}
 \nonumber S &=& \beta^2\frac{\partial F_e}{\partial \beta}\bigg|_{\beta=\beta_e} + \beta^2\frac{\partial F_c}{\partial \beta}\bigg|_{\beta=\beta_c}\\
   &=& \eta_e \frac{A_e}{4} + \eta_c \frac{A_c}{4},\label{entropy}
\end{eqnarray}
with
\begin{eqnarray}
  \eta_e &=& \frac{\delta_e}{90\beta_e\varepsilon_e(\varepsilon_e+\delta_e)} - \frac{\pi n_y^2 \Lambda^{3/2}}{72 \alpha^2 \cos(2\pi/3-\chi)}\ln\frac{\varepsilon_e + \delta_e}{\varepsilon_e};\\
  \eta_c &=& \frac{\delta_c}{90\beta_c\varepsilon_c(\varepsilon_c+\delta_c)} - \frac{\pi n_y^2 \Lambda^{3/2}}{72 \alpha^2 \cos\chi}\ln\frac{\varepsilon_c}{\varepsilon_c + \delta_c},
\end{eqnarray}
where $A_e = 4\pi r_e^2$ and $A_c = 4\pi r_c^2$ are the ares of
two horizons; $\chi$ is determined by effective cosmological
constant, namely, $\chi = \frac{1}{3}\cos^{-1}(-3 M
\sqrt{\Lambda})$. The above simplified representation is obtained
by using the surface gravity expressions
(\ref{Ge})$\sim$(\ref{Go}) and horizons expressions (\ref{re-rc}).
The relationships between inverse Hawking temperature $\beta$ and
horizons are
\begin{eqnarray}
  \beta_e &=& \frac{2\pi}{\kappa_e} = \frac{12\pi}{\Lambda}\cdot\frac{r_e}{(r_c - r_e)(r_e - r_o)};\\
  \beta_c &=& \frac{2\pi}{\kappa_c} = \frac{12\pi}{\Lambda}\cdot\frac{r_c}{(r_c - r_e)(r_c -
  r_o)}.
\end{eqnarray}

It is clear that the 5D black string entropy (\ref{entropy}) is a
linear sum of the area of the black hole horizon and that of the
cosmological horizon. Comparing with 4D case, the 5D Ricci-flat
black string's entropy is modified by the second item of
proportionality coefficients $\eta_e$ and $\eta_c$, which are
discretized by the quantized extra dimensional modes. Furthermore,
it is well known that the small mass approximation in 4D case is
considered as the contribution of the vacuum surrounding black
hole in brick-wall method \cite{wenbiao}, which is very unnatural
and is tried to be solved by many people. However, in this 5D
Ricci-flat black brane solution, the small mass is obtained
naturally by the topological property of far apart binary branes.
When the second brane is sent far away, the ADM mass factor
$\alpha$ tends to infinity. Hence, the effective ADM mass on the
first brane is naturally small. If extra dimension does exist and
is visible near black hole, this nontrivial entropy may can show
something interesting.

\acknowledgments Project supported by the National Basic Research
Program of China (Grant No. 2003CB716300) and National Natural
Science Foundation of China (Grant No. 10573003). We are grateful
to referee's suggestions and also thanks for Feng Luo's help.

\end{document}